\documentclass[prd,amsfonts,noshowpacs,nofootinbib,twocolumn,preprint,10pt]{revtex4}
\usepackage{amsmath,graphicx}

\newcommand{\calR}{{\cal R}}
\begin{document}

\preprint{YITP-10-6}
\title{
A note on the equivalence of a barotropic perfect fluid
with a k-essence scalar field}
\author{Frederico Arroja\footnote{arrojaf{}@{}yukawa.kyoto-u.ac.jp}
 and Misao Sasaki\footnote{misao{}@{}yukawa.kyoto-u.ac.jp}}
\affiliation{Yukawa Institute for Theoretical Physics,
Kyoto University, Kyoto 606-8502, Japan.}

\begin{abstract}
In this brief report, we obtain the necessary and sufficient condition
for a class of noncanonical single scalar field models
to be exactly equivalent to barotropic perfect fluids,
under the assumption of an irrotational fluid flow.
An immediate consequence of this result is that the nonadiabatic
pressure perturbation in this class of scalar field systems vanishes
exactly at all orders in perturbation theory and on all scales.
The Lagrangian for this general class of scalar field models depends
on both the kinetic term and the value of the field.
However, after a field redefinition, it can be effectively cast
in the form of a purely kinetic k-essence model.
\end{abstract}

%\pacs{98.80.-k, 04.30.-w, 98.80.Cq}

\date{February 6, 2010}
\maketitle

%%%%%%%%%%%%%%%%%%%%%%%%%%%%%%%%%%%%%%%%%%%%%%%%%%%%%%%%%%%%%%%%%%%%
%%%%%%%%%%%%%%%%%%%%%%%%%%%%%%%%%%%%%%%%%%%%%%%%%%%%%%%%%%%%%%%%%%%%%
%%%%%%%%%%%%%%%%%%%%%%%%%%%%%%%%%%%%%%%%%%%%%%%%%%%%%%%%%%%%%%%%%%%%%
\section{Introduction}

Both scalar fields and perfect fluids are pervasive in our present models
for the evolution of the Universe since its birth until the current epoch
of accelerated expansion. Both (early time) inflation and the present dark
energy epoch are commonly supposed to be supported by scalar fields.
The intermediate epochs in the evolution of the Universe usually consist
of a radiation-dominated era followed by a matter-dominated era.
These can be well described by barotropic perfect fluids with equations
of state $w$ equal to one third and zero, respectively.
A barotropic perfect fluid can be defined as  a perfect fluid where
the pressure is a function of the energy density
 only (the same function for the background and the perturbations).

Because of the importance of scalar fields and perfect fluids in
present-day cosmological models, the theory of cosmological perturbations
in these models has been intensively studied and is well developed.
For instance, cosmological perturbations for perfect
fluids at linear order have been studied
in \cite{Bardeen:1980kt,Kodama:1985bj,Mukhanov:1990me},
at second order in \cite{Bartolo:2004if,Malik:2008im,Bartolo:2010qu} and
 more recently at third order \cite{D'Amico:2007iw,Christopherson:2009fp}.
Some fully nonlinear results have also been obtained; see for
 example \cite{Lyth:2004gb, Langlois:2005ii}.

Similarly, linear cosmological perturbations in canonical
scalar field systems have been studied in
\cite{Mukhanov:1981xt,Sasaki:1983kd,Sasaki:1986hm,Kodama:1985bj,Mukhanov:1990me},
and in \cite{Maldacena:2002vr,Malik:2006ir,Seery:2006vu,Seery:2008ax}
at second and third order in perturbations. More recently,
cosmological perturbations in noncanonical scalar field models have
attracted much attention. The reason for such interest is that it was
realized that inflationary models in string
theory \cite{Silverstein:2003hf,Chen:2004gc,Chen:2005ad} naturally
contain scalar fields with nonstandard kinetic terms and these models
can also give rise to large non-Gaussianity of the primordial curvature
perturbation. Ref.~\cite{Garriga:1999vw} studied linear perturbations.
For second-order perturbations, see for
instance \cite{Seery:2005wm,Chen:2006nt} and for third-order,
see \cite{Huang:2006eha,Arroja:2008ga,Chen:2009bc,Arroja:2009pd}.
Perturbations up to third order of inflationary models with multiple
noncanonical scalar fields have also been studied; see for
instance \cite{Seery:2005gb,Langlois:2008mn,Langlois:2008wt,Langlois:2008qf,Arroja:2008yy,RenauxPetel:2008gi,Mizuno:2009cv,Mizuno:2009mv,RenauxPetel:2009sj}.

Because of their importance in cosmology, it is natural to ask when,
if in any case, can we describe a scalar field by a perfect fluid
and vice versa. If there are some models where this equivalence is
realized one may be able to use the known results of perturbation
theory mentioned above to study models of perfect fluids or scalar
fields where the calculation in one side of the duality
has not been done. A recent explicit example of such a case can be
found in \cite{Boubekeur:2008kn}. When both results of perturbation
theory for scalar fields and perfect fluids exist, in the dual models,
one can use the equivalence as an extra consistency check for the calculations.

In this brief report, we will discuss the equivalence of a barotropic
perfect fluid with a so-called k-essence/k-inflation scalar
field \cite{ArmendarizPicon:1999rj}.

This paper is organized as follows. In the next section, we will
introduce a general class of k-essence/k-inflation models under
study. We will also present the background equations of motion.
In section \ref{sec:main}, we will briefly discuss linear perturbations;
we shall \emph{obtain} and \emph{solve} the second-order partial
differential equation that gives us the class of scalar field models
that is dual to a barotropic perfect fluid, under the assumption of
an irrotational fluid flow. Finally, section \ref{sec:conclusion}
is devoted to conclusion.

%%%%%%%%%%%%%%%%%%%%%%%%%%%%%%%%%%%%%%%%%%%%%%%%%%%%%%%%%%%%%%%%%%%%%%%
%%%%%%%%%%%%%%%%%%%%%%%%%%%%%%%%%%%%%%%%%%%%%%%%%%%%%%%%%%%%%%%%%%%%%%%%
%%%%%%%%%%%%%%%%%%%%%%%%%%%%%%%%%%%%%%%%%%%%%%%%%%%%%%%%%%%%%%%%%%%%%%%%%
\section{The model\label{sec:model}}

We consider a fairly general class of single scalar
field models described by the action
\begin{equation}
S=\frac{1}{2}\int d^4x\sqrt{-g}\left[M^2_{Pl}R+2P(X,\phi)\right],
\label{action}
\end{equation}
where $\phi$ is the scalar field, $M_{Pl}$ is the Planck mass
that will be set to unity hereafter, $R$  is the Ricci scalar, and
$X\equiv-\frac{1}{2}g^{\mu\nu}\partial_\mu\phi\partial_\nu\phi$ is
the scalar field kinetic term. $g_{\mu\nu}$ is the metric tensor.
We label the scalar field Lagrangian by $P$ and we assume that it is
a well behaved function of two variables, the scalar field $\phi$
 and its kinetic term $X$.

This general field Lagrangian includes as particular cases the
common canonical scalar field model, Dirac-Born-Infeld inflation
\cite{Silverstein:2003hf,Alishahiha:2004eh} and k-inflation/k-essence
\cite{ArmendarizPicon:1999rj}.

The energy-momentum tensor is defined as
\begin{equation}
T_{\mu\nu}=-2\frac{\partial P}{\partial g^{\mu\nu}}+g_{\mu\nu}P\,,
\end{equation}
which gives
\begin{equation}
T_{\mu\nu}=P_{,X}\partial_\mu\phi\partial_\nu\phi+g_{\mu\nu}P\,.
\end{equation}
If $\partial_\mu\phi$ is timelike (i.e. $X>0$), then this is
of the perfect fluid form
\begin{equation}
T_{\mu\nu}=(\rho+P)u_\mu u_\nu+g_{\mu\nu}P\,,
\end{equation}
with the pressure $P$ and the energy density
\begin{equation}
\rho=2XP_{,X}-P\,,
\label{energy}
\end{equation}
where $P_{,X}$ denotes the derivative of $P$ with respect to $X$.
The four-velocity\footnote{This is for a potential flow only.
In general it could have a vector part.} is subject to the
constraint $u_\mu u^\mu=-1$ and reads
\begin{equation}
u_\mu=\frac{\partial_\mu\phi}{\sqrt{2X}}\,.
\end{equation}
The scalar field equation of motion is
\begin{equation}
\nabla^\nu\left(P_{,X}\nabla_\nu\phi\right)+P_{,\phi}=0\,,
\label{fieldeq}
\end{equation}
where $\nabla_\nu$ denotes a covariant derivative.

We are interested in flat, homogeneous, and isotropic
Friedmann-Lema\^{\i}tre-Robertson-Walker background universes
described by the line element
\begin{equation}
ds^2=-dt^2+a^2(t)\delta_{ij}dx^idx^j\,, \label{FRW}
\end{equation}
where $a(t)$ is the scale factor. The Friedmann equation and the
continuity equation read
\begin{equation}
3H^2=\rho_0\,,
\quad
\dot{\rho}_0=-3H\left(\rho_0+P_0\right)\,,
\label{EinsteinEq}
\end{equation}
where the Hubble rate is $H=\dot{a}/a$, and
$\rho_0$ is the background energy density of the scalar field given by
\begin{equation}
\rho_0=2X_0P_{0,X}-P_0\,,
\label{energybackground}
\end{equation}
where $P_{0,X}$ denotes the derivative of $P$ with respect to $X$
evaluated at the background value $X=X_0$.

%%%%%%%%%%%%%%%%%%%%%%%%%%%%%%%%%%%%%%%%%%%%%%%%%%%%%%%%%%%%%%%%%%%%
%%%%%%%%%%%%%%%%%%%%%%%%%%%%%%%%%%%%%%%%%%%%%%%%%%%%%%%%%%%%%%%%%%%%%
%%%%%%%%%%%%%%%%%%%%%%%%%%%%%%%%%%%%%%%%%%%%%%%%%%%%%%%%%%%%%%%%%%%%%
\section{Perturbations and the dual models\label{sec:main}}

In this section, we will first discuss linear perturbations
and the condition for the equivalence between a scalar field
model and a barotropic fluid.
Then we will obtain and solve the second-order partial differential
equation that determines the class of scalar field models dual
 to an irrotational barotropic perfect fluid.
At the end of the section, we will briefly discuss the behavior of
the nonadiabatic pressure perturbation.

On the comoving time-slices, the scalar field fluctuations vanish,
$\delta\phi=0$,
and the three-dimensional spatial metric $h_{ij}$ is perturbed as
\begin{eqnarray}
h_{ij}=a^2\left[(1+2\calR_c)\delta_{ij}
+2\partial_i\partial_jE+2\partial_{(i}F_{j)}\right]\,,
\end{eqnarray}
where tensor perturbations have been neglected and
$\calR_c$ denotes the curvature perturbation on comoving slices.
$E$ and $F_i$ can be put to zero by an appropriate choice of spatial coordinates, where $F_i$ denotes an intrinsic vector perturbation.
The linear equation of motion for $\calR_c$ is
\begin{equation}
\frac{\partial}{\partial t}
\left(\frac{a^3\epsilon}{c_{ph}^2}\,\frac{\partial}{\partial t}\calR_c\right)
-a\,\epsilon\,
\delta^{ij}\frac{\partial^2}{\partial x^i\partial x^j}\calR_c=0\,,
\label{calRceq}
\end{equation}
where $\epsilon=-\dot H/H^2$ and
$c_{ph}$ is the speed of propagation of scalar perturbations
(``speed of sound") given by~\cite{Garriga:1999vw}
\begin{equation}
c_{ph}^2=\frac{P_{0,X}}{\rho_{0,X}}
=\frac{P_{0,X}}{P_{0,X}+2X_0P_{0,XX}}\,.
\label{soundspeed}
\end{equation}

In the case of a barotropic perfect fluid, scalar perturbations
propagate with speed $c_s$ given by
\begin{equation}
c_s^2=\frac{\dot P_0}{\dot \rho_0}\,,
\end{equation}
which is often called the adiabatic sound speed.
The comoving slicing condition for a fluid is $T^0{}_j=0$,
and the curvature perturbation on comoving slices $\calR_c$ follows
Eq.~(\ref{calRceq}) with $c_{ph}^2$ replaced by $c_s^2$.

In general these two speeds are different \cite{Christopherson:2008ry}.
But one can consider under which conditions they are the same.
Requiring $c_{ph}^2=c_s^2$, and after using $\dot P_0=P_{0,X}\dot X+P_{0,\phi}\dot\phi$ and the background equations of motion,
one finds the following equation:\footnote{While this work was being
prepared for publication, the paper~\cite{Unnikrishnan:2010ag}
appeared on the arXiv. They independently found the same equation,
however, they do not solve it in general as we do here and simply
give some very particular solutions of it.}
\begin{equation}
P_{0,\phi}-X_0P_{0,X\phi}+X_0P_{0,\phi}\frac{P_{0,XX}}{P_{0,X}}=0\,.
\label{mastereq}
\end{equation}
This equation can be integrated once to give
\begin{equation}
X_0P_{0,X}=A(\phi)P_{0,\phi}\,,
\label{intonce}
\end{equation}
where $A(\phi)$ is an arbitrary function of $\phi$.
Using the method of characteristics, one can further
integrate Eq.~(\ref{intonce}) to find\footnote{After this work
appeared in the arXiv, we became aware of Ref.~\cite{Akhoury:2008nn}.
They found the same equation (\ref{mastereq}) and the solution (\ref{solution})
for a rather different problem; the necessary condition for the existence
of exact stationary field configurations.
The nontrivial fact that the condition for the existence of
stationary configurations perfectly coincides with
the requirement of exact equivalence between a scalar theory
and a barotropic perfect fluid may deserve further study.}
\begin{equation}
P(X,\phi)=f\bigl(Xg(\phi)\bigr)\,,
\label{solution}
\end{equation}
where $f$ and $g$ are arbitrary functions.
For this Lagrangian, the adiabatic sound speed, which is equal
to the speed of sound, is given by
\begin{equation}
c_{ph}^2=c_s^2=\left(1+2Y\frac{f_{,YY}}{f_{,Y}}\right)^{-1}\,,
\end{equation}
where we define $Y$ as $Y=Xg(\phi)$.
\begin{figure}[t]
\centering
 \scalebox{.5}
 {\rotatebox{0}{
    \includegraphics*{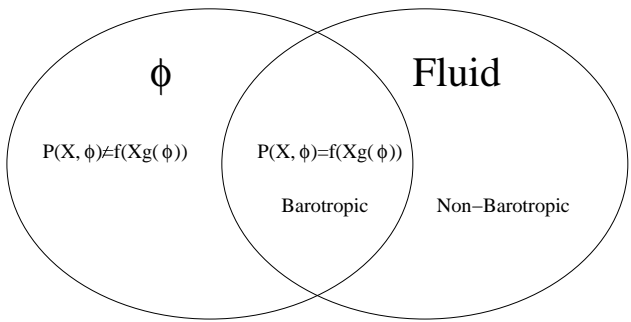}
                 }
 }
\caption{The left ellipse represents the set of all the models
with a general Lagrangian $P(X,\phi)$ while the right ellipse
represents the set of all the perfect fluids. We have shown that
the intersection of these two sets corresponds to barotropic
perfect fluids or scalar field models with Lagrangian
$P(X,\phi)=f(Xg(\phi))$.}\label{Diagram}
\end{figure}
This is the most general class of scalar field models $P(X,\phi)$ that
is exactly equivalent to a barotropic perfect fluid, under the assumption
of the velocity being described by a single scalar potential,
i.e. the fluid flow is irrotational. Figure~\ref{Diagram} is
a graphical description of this result. The fact that any solution
 to the Einstein equations in the presence of a barotropic perfect fluid
can be interpreted as a solution in the presence of a purely kinetic
k-essence scalar field is known in the literature \cite{DiezTejedor:2005fz} (see also for example \cite{Matarrese:1984zw,Quercellini:2007ht}).
 Here we show that all the other k-essence models that depend on the values
of the scalar field and satisfy Eq. (\ref{mastereq}) are also
equivalent to a barotropic perfect fluid. The Lagrangian of these
other models has to be of the form (\ref{solution})\footnote{Ref. \cite{Bertacca:2008uf} showed that the models (\ref{solution}) and the purely kinetic k-essence models possess the same equation of state parameter and speed of sound and thus are also equivalent to barotropic perfect fluids. Here we show that these are the only scalar field models that are equivalent because only these models satisfy the necessary and sufficient condition (\ref{mastereq}).}.

A standard canonic scalar field, i.e. $P(X,\phi)=X-V(\phi)$,
where $V(\phi)$ is the potential, is not included in this class of models.
 The speed of sound is $c_{ph}^2=1$ and the adiabatic sound speed is
\begin{equation}
c_s^2=-1-\frac{\eta-2\epsilon}{3}\,, \quad
\mathrm{with}\quad \epsilon=-\frac{\dot H}{H^2}\,,\quad
 \eta=\frac{\dot \epsilon}{\epsilon H}\,.
\end{equation}
All purely kinetic k-essence models, i.e. the field's Lagrangian is of the form $P(X)$, are included in this class of models. These fields have been used to drive inflation \cite{ArmendarizPicon:1999rj} and to be the dark energy \cite{Copeland:2006wr}.

If $P(X,\phi)=f(Y)$, where $Y=Xg(\phi)$, then Eq.~(\ref{energy})
can be written as
\begin{equation}
\rho=2Yf_{,Y}(Y)-f(Y)\,,
\end{equation}
and one has that the energy density $\rho$ is a function of $Y$ only.
So if $\rho(Y)$ is invertible one can in principle find the equation
 of state $P(\rho)$. Assuming $g>0$ so that $Y>0$,
the previous equation can also be put in the form
\begin{equation}
P(Y)=\frac{1}{2}\sqrt{Y}\int^Y\frac{\rho(Y')}{\sqrt{{Y'}^3}}dY'
+C\sqrt{Y}\,,
\end{equation}
where $C$ is an integration constant.

For a given invertible equation of state one has $\rho+P=F(P)$,
and if $P(X,\phi)=f(Y)$, Eq.~(\ref{energy}) can be written as
\begin{equation}
2f_{,Y}Y=F(f(Y))\,,
\end{equation}
and integrated to give
\begin{equation}
2\int \frac{df}{F(f)}=\ln Y\,,\label{findL}
\end{equation}
from where one finds the Lagrangian given an equation of state $F$.

It may be worth mentioning the expression for the conserved (``baryon'')
number density $n$. For $P=f(Y)$ and $\rho=F(P)-P=2f_{,Y}Y-f(Y)$,
it is determined by
\begin{eqnarray}
\frac{dn}{n}=\frac{d\rho}{\rho+P}
=\frac{f_{,Y}+2f_{,YY}Y}{2f_{,Y}Y}dY\,.
\label{dneq}
\end{eqnarray}
This may be readily integrated to give
\begin{eqnarray}
n=Kf_{,Y}Y^{1/2}\,,
\label{numdensity}
\end{eqnarray}
where $K$ is a constant of integration, which would be a function
of entropy for a perfect fluid.

For the scalar field Lagrangian (\ref{solution}), because both the
pressure and the energy density are functions of one variable only,
i.e. $Y$, it can be shown that the nonadiabatic pressure perturbation
defined for instance in \cite{Wands:2000dp,Malik:2003mv} vanishes
exactly to all orders in perturbations and on all scales.
This is not surprising because all models of the form (\ref{solution})
are dual to barotropic (i.e. adiabatic) perfect fluids where the
nonadiabatic pressure perturbation vanishes by definition.
In other words, the equivalence between the dual models
is independent of the background because the derived second-order
differential equation~(\ref{mastereq}) is independent of the background
but dependent only on the form of $P$ as a function of $X$ and $\phi$,
despite the fact that it was derived from the condition for the equivalence
in linear perturbation theory.

%%%%%%%%%%%%%%%%%%%%%%%%%%%%%%%%%%%%%%%%%%%%%%%%%%%%%%%%%%%%%%%%%%%%%
%%%%%%%%%%%%%%%%%%%%%%%%%%%%%%%%%%%%%%%%%%%%%%%%%%%%%%%%%%%%%%%%%%%%%%
%%%%%%%%%%%%%%%%%%%%%%%%%%%%%%%%%%%%%%%%%%%%%%%%%%%%%%%%%%%%%%%%%%%%
\section{Conclusion\label{sec:conclusion}}

In this brief report, we have studied under which conditions a general
 k-essence scalar field is equivalent to an irrotational barotropic
perfect fluid. We have found that the condition can be written as a
second-order partial differential equation for the Lagrangian of
the field, Eq.~(\ref{mastereq}), that simply states that the sound
speed $c_{ph}$ at which scalar perturbations propagate has to be
 equal to the adiabatic sound speed $c_s$.
We have found the most general solution for that equation,
Eq.~(\ref{solution}). The Lagrangian of the solutions we
found depends explicitly on both the kinetic term $X$ and the scalar
field value $\phi$. However, this dependence is of the restrictive
form $P(X,\phi)=f(Xg(\phi))$, where $f$ and $g$ are arbitrary functions.
After a suitable field redefinition,
$Y=gX=-\frac{1}{2}g^{\mu\nu}\partial_\mu\varphi\partial_\nu\varphi$
where $d\varphi=\sqrt{g}d\phi$,
the Lagrangian can
be cast in the form of a purely kinetic k-essence model.
We note, however, that in reality there may exist plural fields or
fluids that interact with each other, which may lead to small violations
of the perfect fluid/adiabaticity condition. In such a case,
the field redefinition may or may not be useful depending on
the form of interactions as well as on the situation of interest.

One can show that, for the Lagrangian (\ref{solution}), the nonadiabatic
 pressure perturbation vanishes exactly to all others in perturbations
and on all scales. This is because both the pressure and the energy density
are functions of one variable only. This result is less surprising if we
 note that these scalar field models are equivalent to barotropic perfect
fluids where the nonadiabatic pressure perturbation vanishes by definition.

%%%%%%%%%%%%%%%%%%%%%%%%%%%%%%%%%%%%%%%%%%%%%%%%%%%%%%%%%%%%%%%%%%%%%%%
%%%%%%%%%%%%%%%%%%%%%%%%%%%%%%%%%%%%%%%%%%%%%%%%%%%%%%%%%%%%%%%%%%%%%%%%
%%%%%%%%%%%%%%%%%%%%%%%%%%%%%%%%%%%%%%%%%%%%%%%%%%%%%%%%%%%%%%%%%%%%%%%%
\begin{acknowledgments}

We are grateful to David Wands for useful discussions and for
drawing our attention to reference \cite{DiezTejedor:2005fz}.
FA is supported by the Japanese Society for the Promotion of Science
(JSPS). MS is supported in part by JSPS Grant-in-Aid for Scientific
Research (A) No.~21244033, and by JSPS Grant-in-Aid for Creative
Scientific Research No.~19GS0219. This work was also supported in part by MEXT Grant-in-Aid for the global COE program at Kyoto University, "The Next Generation of Physics,
Spun from Universality and Emergence."
\end{acknowledgments}

%%%%%%%%%%%%%%%%%%%%%%%%%%%%%%%%%%%%%%%%%%%%%%%%%%%%%%%%%%%%%%%%%%%%%%
%%%%%%%%%%%%%%%%%%%%%%%%%%%%%%%%%%%%%%%%%%%%%%%%%%%%%%%%%%%%%%%%%%%%%%%
%%%%%%%%%%%%%%%%%%%%%%%%%%%%%%%%%%%%%%%%%%%%%%%%%%%%%%%%%%%%%%%%%%%%%%%%
%\bibliography{bibliography}

\begin{thebibliography}{10}

\bibitem{Bardeen:1980kt}
J.~M. Bardeen,
\newblock Phys. Rev. {\bf D22}, 1882 (1980).
%%CITATION = PHRVA,D22,1882;%%

\bibitem{Kodama:1985bj}
H.~Kodama and M.~Sasaki,
\newblock Prog. Theor. Phys. Suppl. {\bf 78}, 1 (1984).
%%CITATION = PTPSA,78,1;%%

\bibitem{Mukhanov:1990me}
V.~F. Mukhanov, H.~A. Feldman, and R.~H. Brandenberger,
\newblock Phys. Rept. {\bf 215}, 203 (1992).
%%CITATION = PRPLC,215,203;%%

\bibitem{Bartolo:2004if}
N.~Bartolo, E.~Komatsu, S.~Matarrese, and A.~Riotto,
\newblock Phys. Rept. {\bf 402}, 103 (2004), astro-ph/0406398.
%%CITATION = ASTRO-PH/0406398;%%

\bibitem{Malik:2008im}
K.~A. Malik and D.~Wands,
\newblock Phys. Rept. {\bf 475}, 1 (2009), 0809.4944.
%%CITATION = 0809.4944;%%

\bibitem{Bartolo:2010qu}
N.~Bartolo, S.~Matarrese, and A.~Riotto,
\newblock (2010), 1001.3957.
%%CITATION = 1001.3957;%%

\bibitem{D'Amico:2007iw}
G.~D'Amico, N.~Bartolo, S.~Matarrese, and A.~Riotto,
\newblock JCAP {\bf 0801}, 005 (2008), 0707.2894.
%%CITATION = 0707.2894;%%

\bibitem{Christopherson:2009fp}
A.~J. Christopherson and K.~A. Malik,
\newblock JCAP {\bf 0911}, 012 (2009), 0909.0942.
%%CITATION = 0909.0942;%%

\bibitem{Lyth:2004gb}
D.~H. Lyth, K.~A. Malik, and M.~Sasaki,
\newblock JCAP {\bf 0505}, 004 (2005), astro-ph/0411220.
%%CITATION = ASTRO-PH/0411220;%%

\bibitem{Langlois:2005ii}
D.~Langlois and F.~Vernizzi,
\newblock Phys. Rev. Lett. {\bf 95}, 091303 (2005), astro-ph/0503416.
%%CITATION = ASTRO-PH/0503416;%%

\bibitem{Mukhanov:1981xt}
V.~F. Mukhanov and G.~V. Chibisov,
\newblock JETP Lett. {\bf 33}, 532 (1981).
%%CITATION = JTPLA,33,532;%%

\bibitem{Sasaki:1983kd}
M.~Sasaki,
\newblock Prog. Theor. Phys. {\bf 70}, 394 (1983).
%%CITATION = PTPKA,70,394;%%

\bibitem{Sasaki:1986hm}
M.~Sasaki,
\newblock Prog. Theor. Phys. {\bf 76}, 1036 (1986).
%%CITATION = PTPKA,76,1036;%%

\bibitem{Maldacena:2002vr}
J.~M. Maldacena,
\newblock JHEP {\bf 05}, 013 (2003), astro-ph/0210603.
%%CITATION = ASTRO-PH/0210603;%%

\bibitem{Malik:2006ir}
K.~A. Malik,
\newblock JCAP {\bf 0703}, 004 (2007), astro-ph/0610864.
%%CITATION = ASTRO-PH/0610864;%%

\bibitem{Seery:2006vu}
D.~Seery, J.~E. Lidsey, and M.~S. Sloth,
\newblock JCAP {\bf 0701}, 027 (2007), astro-ph/0610210.
%%CITATION = ASTRO-PH/0610210;%%

\bibitem{Seery:2008ax}
D.~Seery, M.~S. Sloth, and F.~Vernizzi,
\newblock JCAP {\bf 0903}, 018 (2009), 0811.3934.
%%CITATION = 0811.3934;%%

\bibitem{Silverstein:2003hf}
E.~Silverstein and D.~Tong,
\newblock Phys. Rev. {\bf D70}, 103505 (2004), hep-th/0310221.
%%CITATION = HEP-TH/0310221;%%

\bibitem{Chen:2004gc}
X.~Chen,
\newblock Phys. Rev. {\bf D71}, 063506 (2005), hep-th/0408084.
%%CITATION = HEP-TH/0408084;%%

\bibitem{Chen:2005ad}
X.~Chen,
\newblock JHEP {\bf 08}, 045 (2005), hep-th/0501184.
%%CITATION = HEP-TH/0501184;%%

\bibitem{Garriga:1999vw}
J.~Garriga and V.~F. Mukhanov,
\newblock Phys. Lett. {\bf B458}, 219 (1999), hep-th/9904176.
%%CITATION = HEP-TH/9904176;%%

\bibitem{Seery:2005wm}
D.~Seery and J.~E. Lidsey,
\newblock JCAP {\bf 0506}, 003 (2005), astro-ph/0503692.
%%CITATION = ASTRO-PH/0503692;%%

\bibitem{Chen:2006nt}
X.~Chen, M.-x. Huang, S.~Kachru, and G.~Shiu,
\newblock JCAP {\bf 0701}, 002 (2007), hep-th/0605045.
%%CITATION = HEP-TH/0605045;%%

\bibitem{Huang:2006eha}
X.~Chen, M.-x. Huang, and G.~Shiu,
\newblock Phys. Rev. {\bf D74}, 121301 (2006), hep-th/0610235.
%%CITATION = HEP-TH/0610235;%%

\bibitem{Arroja:2008ga}
F.~Arroja and K.~Koyama,
\newblock Phys. Rev. {\bf D77}, 083517 (2008), 0802.1167.
%%CITATION = 0802.1167;%%

\bibitem{Chen:2009bc}
X.~Chen, B.~Hu, M.-x. Huang, G.~Shiu, and Y.~Wang,
\newblock JCAP {\bf 0908}, 008 (2009), 0905.3494.
%%CITATION = 0905.3494;%%

\bibitem{Arroja:2009pd}
F.~Arroja, S.~Mizuno, K.~Koyama, and T.~Tanaka,
\newblock Phys. Rev. {\bf D80}, 043527 (2009), 0905.3641.
%%CITATION = 0905.3641;%%

\bibitem{Seery:2005gb}
D.~Seery and J.~E. Lidsey,
\newblock JCAP {\bf 0509}, 011 (2005), astro-ph/0506056.
%%CITATION = ASTRO-PH/0506056;%%

\bibitem{Langlois:2008mn}
D.~Langlois and S.~Renaux-Petel,
\newblock JCAP {\bf 0804}, 017 (2008), 0801.1085.
%%CITATION = 0801.1085;%%

\bibitem{Langlois:2008wt}
D.~Langlois, S.~Renaux-Petel, D.~A. Steer, and T.~Tanaka,
\newblock Phys. Rev. Lett. {\bf 101}, 061301 (2008), 0804.3139.
%%CITATION = 0804.3139;%%

\bibitem{Langlois:2008qf}
D.~Langlois, S.~Renaux-Petel, D.~A. Steer, and T.~Tanaka,
\newblock Phys. Rev. {\bf D78}, 063523 (2008), 0806.0336.
%%CITATION = 0806.0336;%%

\bibitem{Arroja:2008yy}
F.~Arroja, S.~Mizuno, and K.~Koyama,
\newblock JCAP {\bf 0808}, 015 (2008), 0806.0619.
%%CITATION = 0806.0619;%%

\bibitem{RenauxPetel:2008gi}
S.~Renaux-Petel and G.~Tasinato,
\newblock JCAP {\bf 0901}, 012 (2009), 0810.2405.
%%CITATION = 0810.2405;%%

\bibitem{Mizuno:2009cv}
S.~Mizuno, F.~Arroja, K.~Koyama, and T.~Tanaka,
\newblock Phys. Rev. {\bf D80}, 023530 (2009), 0905.4557.
%%CITATION = 0905.4557;%%

\bibitem{Mizuno:2009mv}
S.~Mizuno, F.~Arroja, and K.~Koyama,
\newblock Phys. Rev. {\bf D80}, 083517 (2009), 0907.2439.
%%CITATION = 0907.2439;%%

\bibitem{RenauxPetel:2009sj}
S.~Renaux-Petel,
\newblock JCAP {\bf 0910}, 012 (2009), 0907.2476.
%%CITATION = 0907.2476;%%

\bibitem{Boubekeur:2008kn}
L.~Boubekeur, P.~Creminelli, J.~Norena, and F.~Vernizzi,
\newblock JCAP {\bf 0808}, 028 (2008), 0806.1016.
%%CITATION = 0806.1016;%%

\bibitem{ArmendarizPicon:1999rj}
C.~Armendariz-Picon, T.~Damour, and V.~F. Mukhanov,
\newblock Phys. Lett. {\bf B458}, 209 (1999), hep-th/9904075.
%%CITATION = HEP-TH/9904075;%%

\bibitem{Alishahiha:2004eh}
M.~Alishahiha, E.~Silverstein, and D.~Tong,
\newblock Phys. Rev. {\bf D70}, 123505 (2004), hep-th/0404084.
%%CITATION = HEP-TH/0404084;%%

\bibitem{Christopherson:2008ry}
A.~J. Christopherson and K.~A. Malik,
\newblock Phys. Lett. {\bf B675}, 159 (2009), 0809.3518.
%%CITATION = 0809.3518;%%

\bibitem{Unnikrishnan:2010ag}
S.~Unnikrishnan and L.~Sriramkumar,
\newblock (2010), 1002.0820.
%%CITATION = 1002.0820;%%

\bibitem{Akhoury:2008nn}
R.~Akhoury, C.~S. Gauthier, and A.~Vikman,
\newblock JHEP {\bf 03}, 082 (2009), 0811.1620.
%%CITATION = 0811.1620;%%

\bibitem{DiezTejedor:2005fz}
A.~Diez-Tejedor and A.~Feinstein,
\newblock Int. J. Mod. Phys. {\bf D14}, 1561 (2005), gr-qc/0501101.
%%CITATION = GR-QC/0501101;%%

\bibitem{Matarrese:1984zw}
S.~Matarrese,
\newblock Proc. Roy. Soc. Lond. {\bf A401}, 53 (1985).
%%CITATION = PRSLA,A401,53;%%

\bibitem{Quercellini:2007ht}
C.~Quercellini, M.~Bruni, and A.~Balbi,
\newblock Class. Quant. Grav. {\bf 24}, 5413 (2007), 0706.3667.
%%CITATION = 0706.3667;%%

\bibitem{Bertacca:2008uf}
D.~Bertacca, N.~Bartolo, A.~Diaferio, and S.~Matarrese,
\newblock JCAP {\bf 0810}, 023 (2008), 0807.1020.
%%CITATION = 0807.1020;%%

\bibitem{Copeland:2006wr}
E.~J. Copeland, M.~Sami, and S.~Tsujikawa,
\newblock Int. J. Mod. Phys. {\bf D15}, 1753 (2006), hep-th/0603057.
%%CITATION = HEP-TH/0603057;%%

\bibitem{Wands:2000dp}
D.~Wands, K.~A. Malik, D.~H. Lyth, and A.~R. Liddle,
\newblock Phys. Rev. {\bf D62}, 043527 (2000), astro-ph/0003278.
%%CITATION = ASTRO-PH/0003278;%%

\bibitem{Malik:2003mv}
K.~A. Malik and D.~Wands,
\newblock Class. Quant. Grav. {\bf 21}, L65 (2004), astro-ph/0307055.
%%CITATION = ASTRO-PH/0307055;%%

\end{thebibliography}
%\bibliographystyle{bibstylefile}

\end{document}